**Quantum Batteries: A Materials Science Perspective**

*Andrea Camposeo,[1] Tersilla Virgili,[2] Floriana Lombardi,[3] Giulio Cerullo,[2,4] Dario Pisignano,[1,5] Marco Polini[5]*

[1]NEST, Istituto Nanoscienze - CNR and Scuola Normale Superiore, Piazza San Silvestro 12, I-56127 Pisa, Italy

[2]Istituto di Fotonica e Nanotecnologie - CNR, IFN, 20133 Milano, Italy

[3]Department of Microtechnology and Nanoscience, Chalmers University of Technology, SE-41296 Göteborg, Sweden

[4]Dipartimento di Fisica, Politecnico di Milano, Piazza Leonardo da Vinci 32, Milano 20133, Italy

[5]Dipartimento di Fisica "E. Fermi", Università di Pisa, Largo B. Pontecorvo 3, I-56127 Pisa, Italy, E-mail: dario.pisignano@unipi.it

**Abstract**.

In the context of quantum thermodynamics, quantum batteries have emerged as promising devices for energy storage and manipulation. Over the past decade, substantial progress has been made in understanding the fundamental properties of quantum batteries, with several experimental implementations showing great promise. This Perspective provides an overview of the solid-state materials platforms that could lead to fully operational quantum batteries. After briefly introducing the basic features of quantum batteries, we discuss organic microcavities, where superextensive charging has already been demonstrated experimentally. We then explore other materials, including inorganic nanostructures (such as quantum wells and dots), perovskite systems, and (normal and high-temperature) superconductors. Key achievements in these areas, relevant to the experimental realization of quantum batteries, are highlighted. We also address challenges and future research directions. Despite their enormous potential for energy storage devices, research into advanced materials for quantum batteries is still in its infancy. This paper aims to stimulate interdisciplinarity and convergence among different materials science research communities to accelerate the development of new materials and device architectures for quantum batteries.

# 1. Introduction

The quest for innovative materials and device architectures for energy storage is a constantly advancing scientific and technological domain that greatly influences our daily lives. The development of Li-ion batteries has contributed significantly to the portable electronics market,[1, 2] while the growing number of electric vehicles and green energy production demand novel materials to enhance the energy density, lifetime, and safety of batteries.[3-5] Various innovative materials and processes are being investigated to improve the properties of current batteries.[4-7]

Nowadays, the advancement of quantum technologies such as quantum information, computation, simulation, and sensing calls for novel and specific platforms for energy storage. Quantum devices require solutions for energy storage and energy provider components that should be highly compatible with their unique architectures, and capable of supporting the creation of a superposition of quantum states. The need for specialized energy storage units for reversible quantum operations has been recently highlighted.[8]

In this context, quantum batteries (QBs), first introduced by Alicki and Fannes in 2013,[9] have emerged as an intriguing approach to energy storage.[10,11] QBs are devices that are made of quantum systems, and they harness fundamental quantum mechanical effects to charge, store, and release energy. QBs can be charged/discharged through operations that establish coherent superpositions among various states,[12] while the energy is stored in the excited states of the quantum systems. Entanglement effects and non-classical correlations have been shown to lead to superextensive scaling of the charging power (that is, the ratio between the stored energy and the charging time), with a quantum advantage compared to their classical counterparts.[10]

The first model of a QB amenable to laboratory fabrication, proposed by Ferraro *et al*.,[13] consisted of a solid-state device obtained by coupling $N$ two-level systems (TLSs) to the optical mode of a cavity. The charging power of this device, which was dubbed "Dicke QB", was shown to display superlinear scaling, with a $\sqrt{N}$ improvement resulting from the collective behavior mediated by the cavity field. This opens interesting perspectives for energy storage devices that can be charged faster.

To date, QBs have primarily been studied from a theoretical viewpoint, evidencing many unique properties.[11 and references therein] While various experimental designs for quantum batteries have been proposed,[11, 14, 15] only a few preliminary, yet significant steps toward the practical demonstration of a fully operational QB have been reported.[16-19] These steps include demonstrating superabsorption properties in microcavities with organic molecules using ultrafast pump-probe spectroscopy,[16] energy storage in a single-spin QB for up to two minutes,[18] and the experimental investigation of charging (discharging) and storage capabilities of QBs based on single qutrit.[17,19]

This Perspective provides an overview of the most promising platforms for the experimental realization of QBs, focusing on the advanced materials and device architectures that can be potentially used. After introducing the main properties of QBs and the different schemes proposed for their implementation, we present an overview of the materials that are potentially suitable for the realization of some QB architectures. First, we review microcavities with organic molecules and inorganic nanostructures, such as quantum wells (QWs) and quantum dots (QDs), highlighting examples where either strong light-matter coupling or collective effects are demonstrated. We then discuss other systems, including microcavities with perovskite materials and superconducting circuits. Motivated by theoretical predictions of a genuine quantum advantage in QBs based on the Sachdev-Ye-Kitaev (SYK) model,[20] we briefly discuss strange metals as potential materials for the implementation of these exotic batteries. Other quantum coherent systems such as trapped ions, Rydberg atoms, and ultracold atoms, for which recent Reviews and Perspective articles are already available,[21-27] are not discussed in this Perspective. Finally, we discuss some critical aspects of achieving a fully operational QB, and some possible and promising future research directions.

## 2. Quantum Batteries

A QB is a quantum mechanical system with a non-zero energy bandwidth whose Hamiltonian, $\mathcal{H}_0$, can be written as a sum of $N$ local terms, each described by a microscopic Hamiltonian $h_i$:

$$\mathcal{H}_0 = \sum_{i=1}^{N} h_i. \qquad (1)$$

An ensemble of TLSs, for example, is the simplest form of a QB one can think of.[13] In this case, $h_i$ physically represents the microscopic Hamiltonian of each TLS. Note that, in contrast to a classical battery, a QB can be acted upon via unitary operations that may create, at least temporarily, nonclassical correlations between the $N$ subunits.

As in the case of any other battery, three operational stages can be defined: a charging stage, a storage stage, and a discharging stage (where work is extracted from the battery). In the charging stage of the non-equilibrium dynamics, one switches on an interaction $\mathcal{H}_1$ in a time window of duration $\tau_c$, in such a way to inject energy into the battery. This can be done provided that the commutator $[\mathcal{H}_0, \mathcal{H}_1] \neq 0$, a condition that is essential for charging the actual QB, which is mathematically described the Hamiltonian $\mathcal{H}_0$. Under the action of $\mathcal{H}_1$, the state $\rho(t)$ of the battery changes according to the time-dependent Schroedinger equation. Real devices, of course, are always open quantum systems, and therefore this closed-system Hamiltonian description needs to be transcended in modeling actual laboratory effects, by using the theory of open quantum systems. In this respect, theoretical results have been recently reported by considering states that are protected from the environment.[28, 29] A more detailed discussion of open QBs can be found in Ref. [11] and the references cited therein.

An optimal time $\tau^*$ can be defined in such a way that a particular figure of merit of the battery is maximized. For example, $\tau^*$ can be defined as the time at which the energy stored in the battery degrees of freedom, $E(t) = Tr[\rho(t)\mathcal{H}_0] - Tr[\rho_0\mathcal{H}_0]$, is maximized, where $\rho_0$ is the initial state of the battery. Similar optimally-defined time scales can be associated to other figures of merit, such as the instantaneous power $P_{inst}(t) = dE(t)/dt$ or the average power $P(t) = \frac{E(t)}{t}$. We finally mention another important figure of merit, which is the ergotropy,[30] physically measuring the maximum amount of work that can be extracted from a QB in a cyclic process such that the Hamiltonian of the system is the same in the initial and final states. Moreover, the charging/discharging phases of QBs are dynamic processes which can be specifically controlled to enhance the device figures of merit. To this aim, various approaches have been proposed which include the exploitation of nonreciprocity through reservoir engineering,[31] squeezing effects,[32] virtual photons,[33] catalysis,[34] shortcuts to adiabaticity,[35] and optimal quantum control[36] among the others. More details can be found in Ref. [11] and the references cited therein.

As stated above, a particular implementation of a quantum energy storage device is the Dicke QB.[13] The name "Dicke" was attached to this proposal precisely because this QB is composed of $N$ identical TLSs coupled to the very same photonic mode of a cavity. The Dicke model,[37] indeed, describes the interaction between an ensemble of emitters and a common radiation field. These systems feature interesting properties such as superradiance,[38] which is a collective phenomenon where an ensemble of interacting excited emitters decays much faster than one in which interactions can be neglected. This results in an enhanced radiative decay rate and a superextensive scaling of emission intensity, growing with the number of emitters as $N^2$ rather than $N$. Superradiance, originally demonstrated in the context of atomic and molecular physics,[38] has been observed in various solid-state systems, such as molecular

H-aggregates,[39] QDs,[40] organic single crystals,[41] single diamond nanocrystals at room temperature,[42] and color centers in diamond.[43]

For practical implementations of a solid-state Dicke QB, TLSs can be realized by using any material that enables the realization of a quantum system with a discrete energy spectrum containing a ground and an excited state, well separated by the rest of the spectrum. Examples include (but are not limited to) QDs, spin systems such as nitrogen-vacancy (NV) centers in diamond, and superconducting qubits such as transmons. Interest in this type of QBs is mainly related to the potential ease of experimental fabrication in solid-state devices, a recent example being that of organic molecules in Fabry-Pérot cavities.[16]

Ferraro *et al.* calculated the maximum charging power for $N$ identical batteries, each composed of a TLS coupled to a cavity mode and operated in parallel (Figure 1a), as well as for a Dicke QB with $N$ TLS (Figure 1b).[13] The latter configuration, termed "collective charging" has a clear advantage in exhibiting a charging power higher by a factor of $\sqrt{N}$ compared to the parallel charging setup. Further studies[44] revealed that the enhanced charging power of the Dicke QB originates from many-body collective effects rather than purely quantum effects.

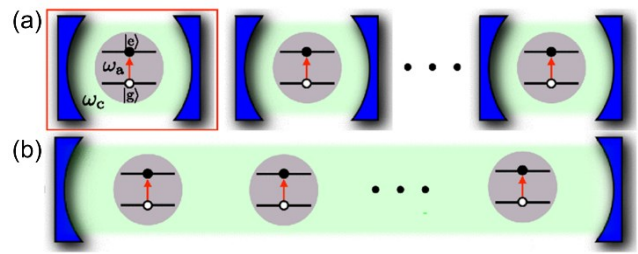

**Figure 1**. (a) Schematic illustration of an array of QBs operating in parallel. Each battery is made of a TLS (energy separation between the ground and the excited state: $\hbar\omega_a$) coupled to a separate photonic cavity (blue). The red arrow indicates a transition from the ground state to the excited state induced by the photon field. (b) Illustration of a Dicke QB, for which an ensemble of TLSs is embedded in the same cavity and interacts with the same photonic mode. Reproduced with permission.[13] Copyright 2018, American Physical Society.

Other physical systems have been designed and theoretically studied for the implementation of a QB.[11 and references therein] One example is that of ensembles of interacting spins. Interacting spin systems have been investigated in the framework of condensed matter physics, quantum communication, and computation. At variance with Dicke QBs, direct spin-spin interactions play a pivotal role in spin batteries (in a Dicke QB direct interactions between TLSs are neglected, while effective interactions between TLS are solely mediated by the common cavity mode).

Recently, systems based on single photons have been used for the experimental simulation of various charging protocols of QBs, with the aim of investigating the role of coherence and entanglement,[45] the effect of a catalyst interposed between the charger and QB,[46] and the charging by indefinite casual order.[47]

A genuine (rather than collective) quantum advantage was demonstrated by Rossini *et al.*[20] who introduced a model of a QB based on TLSs interacting via the SYK Hamiltonian.[48-50] The SYK model describes interactions between constituents of matter that are so strong to defy the usual paradigm we use to interpret the behavior of interacting electrons in metals and semiconductors, i.e. Landau theory of normal Fermi liquids. Current proposals to implement the SYK model rely on cold atoms,[51] topological superconductors,[52, 53] and graphene QDs with irregular boundaries in strong applied magnetic fields.[54-56] Furthermore, spatially random interactions, which are at the core of the SYK model, are believed to be responsible for the formation of an exotic phase of matter dubbed "strange metal",[57] which exists in high-temperature superconductors[58] and twisted bilayer graphene at the magic angle.[59] We, therefore, anticipate that these two material platforms may play a role in future experimental implementations of QBs.

In addition to QBs based on the SYK model, Andolina *et al.*[60] have recently proposed a novel QB that features a genuine quantum advantage. Here, the charger and the battery are harmonic oscillators coupled by non-linear interaction during the charging phase. Such non-linear bosonic quantum batteries might be realized by using superconducting circuits.

## 3. Advanced Materials for Quantum Batteries

For the experimental implementation of QBs, several characteristics of the used materials and device architecture should be carefully evaluated. A critical feature is the energy difference between the levels of the considered materials (i.e. the level spacing), $\Delta E_{TLS}$. Values significantly exceeding thermal energy at room temperature (about 25 meV) are less sensitive to relaxation phenomena. Similarly, in excitonic systems, the exciton binding energy should be higher than the thermal energy for them to remain stable at room temperature. For Dicke QBs, a high quality factor ($Q$) of the cavity is also relevant to enhance the interaction between the TLSs and the electromagnetic field of the microcavity, as well as the tunability of the cavity and material properties by external fields in order to implement different charging/discharging protocols. Other critical features are the capability of storing the energy for long time intervals and the availability of material processing technologies that can be easily scaled up.

Different material platforms have been proposed for the experimental implementation of QBs.[11, 14, 15] In this section, we review the most promising ones and discuss how features that are critical for QB implementation can be addressed. After presenting the properties and structures of proposed Fabry-Pérot microcavities, Section 3.1 delves into organic molecules used to study exciton-photon coupling in microcavities. Section 3.2 and 3.3 report on potential inorganic semiconductors and perovskite materials coupled to Fabry-Pérot microcavities, emphasizing collective phenomena and exciton-photon coupling that might be useful for QB design and fabrication. Section 3.4 highlights superconductors circuits with single and few qubits coupled to microwave microcavities. Section 3.5 introduces materials that can be used as spin arrays for the realization of QBs. Lastly, Section 3.6 discusses the properties of strange metals as a potential platform for the implementation of QBs that display a genuine quantum advantage. For information on other systems currently investigated in quantum technologies and potentially usable for the realization of QBs, such as trapped ions, Rydberg atoms, ultracold atoms and Bose-Einstein condensates, the readers are referred to recent comprehensive viewpoints and reviews.[21-27]

## 3.1. Organic Microcavities

One of the platforms for the implementation of QBs relies on microcavities enclosing an ensemble of organic molecules. Here, the Fabry-Pérot resonator is typically used as the microcavity architecture. It is formed by a layer of an organic material sandwiched between two high-reflectivity plane parallel mirrors. Mirrors can be thin metallic films, distributed Bragg reflectors (DBRs) or combinations of them (Figure 2a). DBRs are one-dimensional photonic crystals consisting of a stack of $N_p$ pairs of alternating layers of a high ($n_H$) and low ($n_L$) refractive index material with minimal absorption, in which constructive interference among multiple reflections generates a broad band with high reflectivity (stopband) for light propagating along the stack direction (Figure 2b-d). The highest reflectivity at the central wavelength of the stopband ($\lambda_{SB}$) is achieved in the so-called $\lambda/4$ configuration. This scheme is implemented with thicknesses of the high ($t_H$) and low ($t_L$) refractive index materials: $t_H = \lambda_{SB}/4n_H$ and $t_L = \lambda_{SB}/4n_L$, respectively. Typical pairs of materials employed for DBRs in organic microcavities are $SiO_2/TiO_2$,[61] $SiO_2/Ta_2O_5$,[62, 63] $SiO_2/HfO_2$,[64] $SiO_2/Nb_2O_5$,[65] $TeO_x/LiF$,[66] $SiO_2/SiN_x$,[67, 68] $MgF_2/ZnS$,[68] Zircone/$ZrO_2$,[69] $CaF_2/ZnS$,[70] and $AlAs/Al_{0.5}GaAs$.[71] DBRs can be fabricated by different deposition technologies, including either physical or chemical vapor deposition, electron beam evaporation, molecular beam epitaxy (MBE), and sputter deposition. It is worth noting that the reflectivity of DBRs can be negatively affected by the bulk and surface inhomogeneities of the deposited layers. By properly engineering the deposition processes, layers with root mean square roughness <2 nm can be deposited and reflectivity exceeding 99% is achieved for $N_p > 10$.[61, 66] Recently, McGhee et al.[72] introduced hybrid metal-DBR mirrors made of a thick Ag layer coated with a few $SiO_2$/$Nb_2O_5$ $\lambda$/4 pairs (Figure 2e,f). These hybrid mirrors allow for achieving broadband reflectivity and enhanced confinement, as well as simplifying the fabrication method.

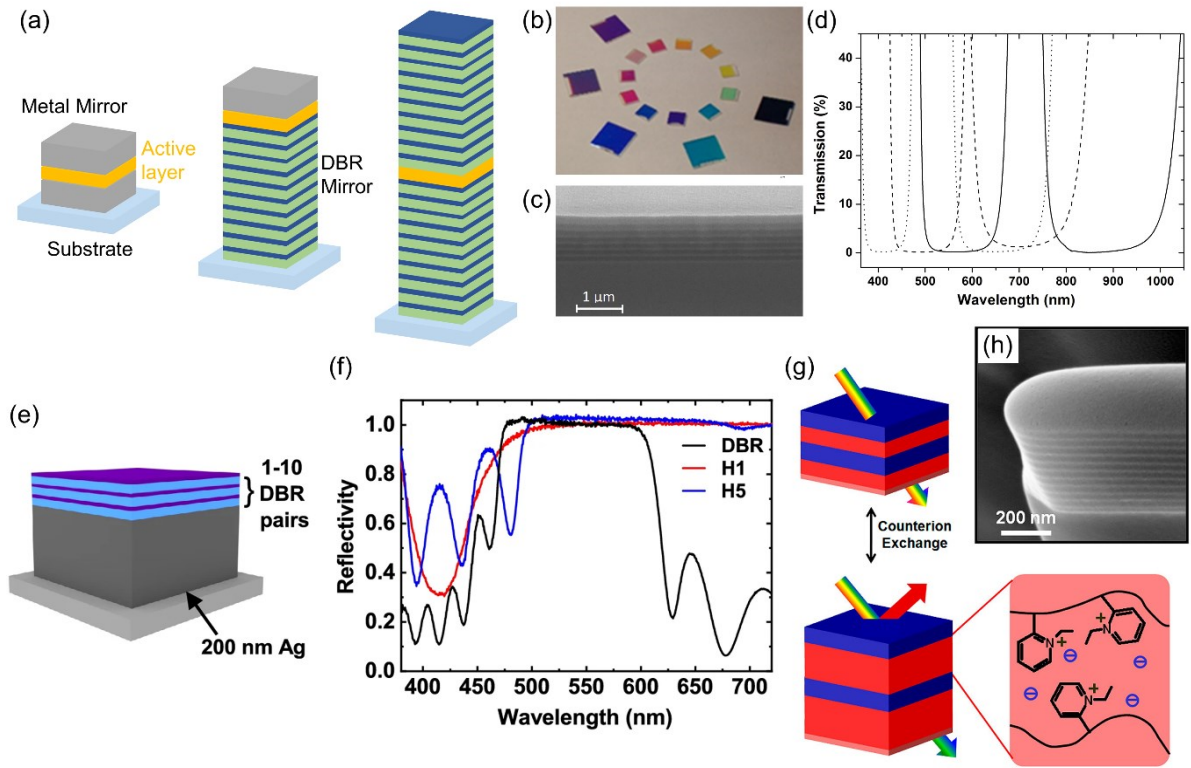

**Figure 2**. (a) Schematic illustrations of different Fabry-Pérot microcavities with either metal or DBR mirrors, as well as a combination of them. (b) Photograph of different DBR mirrors with stopbands in the visible and near-infrared made by reactive electron-beam deposition. (c) Cross-sectional scanning electron micrograph of a $SiO_2$/$TiO_2$ DBR. (d) Example of transmission spectra of the DBRs, with $\lambda_{SB}$ in the interval 425-850 nm. (b)-(d) Reproduced with permission.[61] Copyright 2006, Optical Society of America. (e) Schematic illustration of a hybrid metal/DBR mirror. (f) Reflectivity spectra of a 10-pair DBR mirror, and hybrid metal/DBR mirrors with 1 (H1) and 5 (H5) dielectric pairs, respectively. (e)-(f) Reproduced under terms of the CC-BY license.[72] Copyright 2021, The Authors, published by Springer Nature. (g) Illustration of the mechanism of stopband tuning by counterions exchange in lamellar films. (h) Cross-sectional scanning electron micrograph of a dry lamellar film. (g)-(h) Reproduced with permission.[75] Copyright 2012, American Chemical Society.

DBRs can also be manufactured by alternating polymer and nanocomposite layers with different refractive indexes.[73, 74] These DBRs can achieve reflectivity > 95% and be processed by solution-based methods such as spin-coating, dip coating, and doctor blading.[74] Moreover, polymeric materials with refractive indexes variable by external physical and chemical stimuli can be used for DBR fabrication (Figure 2g,h), with the possibility of finely tuning the spectral features of the stopband.[75]

The allowed modes of the microcavity can be determined by considering it as a Fabry-Pérot interferometer (etalon) with resonant wavelengths $\lambda_R$ given by: $m\lambda_R = 2nd\,cos(\theta)$, where $m$ is the cavity mode order, $n$ and $d$ are the refractive index and the thickness of the active layer, respectively, and $\theta$ is the incidence angle of light. The microcavity is characterized by the $Q$ factor, which describes the decay of energy in the cavity and can be defined as: $Q = 2\pi E_{sto}/E_{dis}$, where $E_{sto}$ is the energy stored in the resonator and $E_{dis}$ is the energy dissipated per oscillation cycle. The $Q$ factor can also be written as the ratio $Q = \nu_R/\Delta\nu$, where $\nu_R$ is the resonance frequency of the cavity and $\Delta\nu$ is the mode linewidth. The $Q$ factor is in turn related to the photon lifetime in the cavity, which is the time for which light remains trapped between the cavity mirrors and can be expressed as: $\tau_{cav} = \frac{1}{\kappa_{cav}} = \frac{Q}{2\pi\nu_R}$, where $\kappa_{cav}$ is the cavity photon decay rate. Cavity quality factors can be improved by using dielectric mirrors and uniform interfaces between the various layers of the device.

In organic microcavities, the active layer is made of molecular semiconductors, in which photoexcitation creates excitons. The binding energy of the exciton and thus its effective radius, depends on the screening of the Coulomb force by the medium surrounding the charges. For molecular semiconductors, weak screening due to the low dielectric constant of the medium results in tightly bound excitons with small radius, the so-called Frenkel excitons. Typical exciton linewidths in organic semiconductors are on the order of $10^{-1}$-1 eV.[76, 77] The linewidth is determined by the broadening of the transition, by the presence of a vibronic progression, and by intermolecular interactions. In particular, broadening mechanisms of a transition can be distinguished as homogeneous and inhomogeneous. In a homogeneously broadened transition, all the molecules in the ensemble experience the same interaction with the environment and have the same average transition energy. The homogeneous linewidth of a transition, $\Delta\nu_{hom}$, is inversely proportional to the dephasing time $T_2$ (the decay time for the electronic coherence between the ground and excited state produced by the photonic excitation) according to the expression: $\Delta\nu_{hom} = 1/(\pi T_2)$. To ensure a narrow excitonic linewidth and consequently a longer dephasing time, it is necessary to use very rigid molecules, such as porphyrins or dyes like cyanine.[78] For cyanines, intermolecular interactions can lead to the so-called J-aggregates, head-to-tail arrangements of molecules in which excitonic coupling among the lowest energy optical transitions leads to intense and red-shifted absorption and emission bands.[79, 80] In J-aggregates coupling to intramolecular vibronic modes is reduced, resulting in small Stokes shifts and linewidths which can be as small as tens of meV.[78] Other materials, like the porphyrins[81] and BODIPY-based molecules,[82] have similar characteristics when dispersed in polymer matrices, but with broader excitonic linewidth of ≈100 meV.

In addition, the microcavities incorporating organic materials can operate in different regimes (weak, strong, and ultrastrong coupling), depending on the exciton-photon coupling rate, $g$. When considering the exciton decay rate, $\kappa_{exc}$, the weak coupling regime corresponds to the condition $g < (\kappa_{exc}, \kappa_{cav})$. In this regime, either suppression or enhancement of the spectral and spatial characteristics of the excitonic spontaneous emission might occur. In particular, if the exciton is resonant with the cavity mode, then the radiative decay rate is enhanced (Purcell effect),[83] while radiative decay is suppressed if the exciton is off resonance with the cavity mode.

On the other hand, if $g > (\kappa_{exc}, \kappa_{cav})$ the cavity is in the strong coupling regime. In this case, the rate at which excitons and cavity modes exchange energy exceeds the rate at which either excitons or cavity modes decay. The exciton and the cavity mode become two coupled oscillators that exchange energy with a frequency $\Omega = 2g$ (Rabi frequency). In the strong coupling regime, new quasiparticles are formed called exciton-polaritons, a coherent superposition of exciton and photon states. Two polariton branches are formed, which are called the lower polariton branch (LPB) and upper polariton branch (UPB), at energies lower and higher, respectively, compared to the excitonic transition. The polariton branches undergo anticrossing around the condition in which excitons and cavity modes are energetically degenerate. At zero cavity-transition detuning, the polariton branches are separated in energy by the Rabi splitting. To achieve the strong coupling regime, one needs to have preferably a high $Q$ factor cavity and an exciton transition with a large oscillator strength and a narrow linewidth (comparable with the one of the cavity mode).

In the strong-coupling regime, the excitons in the cavity absorb and spontaneously re-emit a photon many times before dissipation becomes effective, giving rise to mixed light-matter eigenmodes.[84] In some cases, photon exchange may occur on timescales comparable to the oscillation period of light, leading to coupling strengths comparable to the transition frequencies of a system. This regime is called ultrastrong coupling.[85, 86]

Lidzey et al.[81] reported, for the first time, an organic semiconductor microcavity that operates in the strong-coupling regime. They used as organic semiconductor the tetra-(2,6-t-butyl)phenol-porphyrin zinc (4TBPPZn) dispersed in a polystyrene (PS) matrix (Figure 3a). The organic layer was placed between a silver mirror and a DBR mirror, the latter consisting of nine alternating pairs of silicon nitride and silicon dioxide. Typical $Q$ factors for these structures were $\approx 125$, and they achieved a Rabi splitting of around 160 meV. Different cavities were fabricated using as active layer cyanine dye J-aggregates[78] and bromine-substituted boron-dipyrromethene (BODIPY-Br)[87] (Figure 3b,c). For cyanine dye J-aggregates the organic layer was sandwiched between a silver and a DBR mirror, reaching a $Q$ factor of 85 and a room temperature Rabi splitting of around 80 meV.[78] For the BODIPY-based molecules, these were dispersed in PS and spin-cast onto a DBR consisting of 10 pairs of $SiO_2/Nb_2O_5$. Then, on top of the PS film a second 8-pair $SiO_2/Nb_2O_5$ DBR film was deposited by using ion assisted electron beam and reactive sublimation. The measured $Q$ factor of the resulting cavity was $\approx 440$ and a Rabi splitting of approximately 91 meV was achieved.

A different approach to achieve a strong coupling regime is based on the coupling of the cavity photon with one or more vibronic replicas of the exciton molecular transition. In this case molecular films of crystalline anthracene,[67] oligofluorenes,[88] methyl-substituted ladder-type poly(*p*-phenylene) (MeLPPP)[62] and fluorescent protein[89] (Figure 2d-f) were fabricated. For the green fluorescent protein, both the absorption and emission spectrum are significantly broadened and show considerable overlap, in contrast to all the effort that has been directed toward finding organic materials with particularly narrow linewidths so that the polariton linewidth is smaller than the Rabi splitting of the coupled system. For these microcavities, a different approach was followed: by increasing the overall cavity thickness and coupling several cavity modes to the excitonic transitions of the protein, the theoretical $Q$ factor of the involved cavity modes increases by up to 50,000 and increases the photonic character of the coupled cavity polaritons (86% photonic), which then also reduces the polariton linewidth.

Ultrastrong exciton–photon coupling of Frenkel molecular excitons was also demonstrated at room temperature in a metal-clad microcavity ($Q\sim 30$) containing a thin film of 2,7-bis[9,9-di(4-methylphenyl)-fluoren-2-yl]-9,9-di(4-methylphenyl)fluorene (TDAF).[90] A giant Rabi splitting $\sim 1$ eV was achieved. Reversible optical switching from the weak to the ultrastrong coupling regime was demonstrated in microcavity incorporating photochromic molecules, with Rabi splitting up to 700 meV.[91]

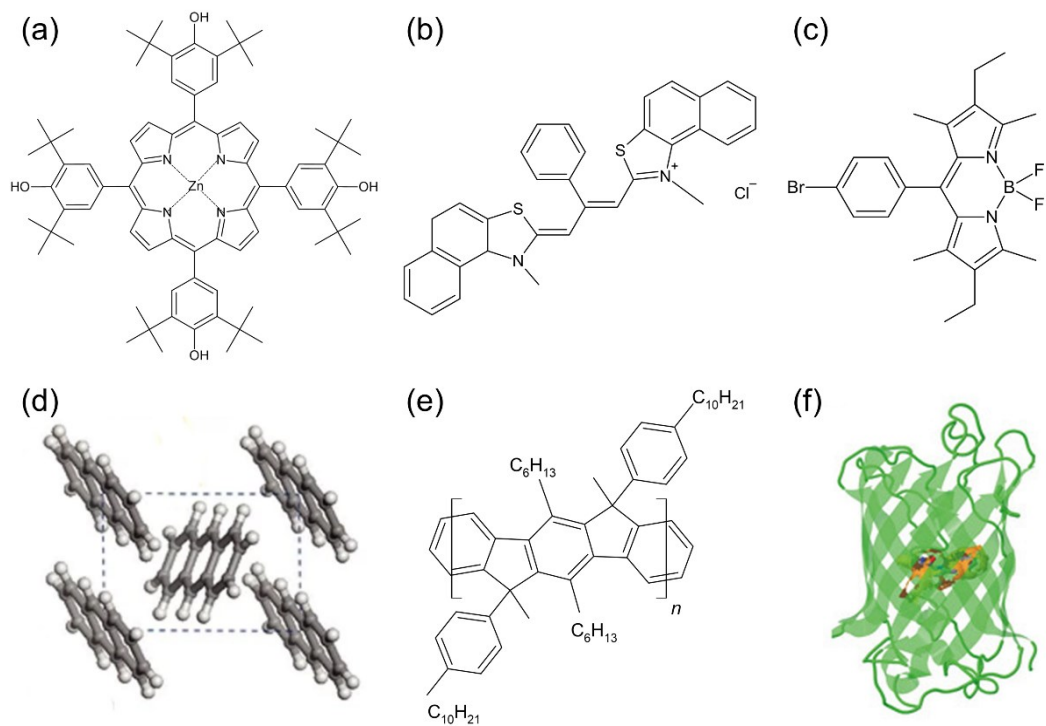

**Figure 3**: Chemical structures of examples of organic molecules and biomolecules used for investigating strong coupling effects in microcavities. (a) 4TBPPZn; (b) 2, 2´-dimethyl-8-phenyl-5, 6, 5´, 6´-dibenzothiacarbocyanine chloride. (c) BODIPY-Br. (d) Illustration of the anthracene crystal structure. Reproduced with permission.[67] Copyright 2010, Springer Nature Limited. (e) MeLPPP. (f) Enhanced green fluorescent protein. Reproduced under terms of the CC-BY license.[89] Copyright 2016, The Authors, published by American Association for the Advancement of Science.

These examples highlight the rich variety of organic materials that could be utilized for the realization of a Dicke QB. This has been recently explored by Quach *et al.*,[16] who experimentally demonstrated the concept of Dicke QB using a microcavity in which the active material consists of organic molecules, Lumogen-F Orange (LFO), dispersed in a transparent PS matrix (Figure 4a). The LFO absorption and photoluminescence (PL) spectra are shown in Figure 4b. By operating in the vicinity of the 0-0 electronic transition, which is the transition between the lowest vibrational states of the ground and excited electronic states, the LFO molecules can be regarded as TLSs. The number of coupled TLS was controlled by regulating the concentrations of the dye molecules. The charging and energy storage dynamics were characterized by ultrafast transient absorption (TA) spectroscopy. In TA spectroscopy, the sample under study (in this case a cavity enclosing the dye molecules) is excited by an ultrashort pump pulse, and the energy stored in the sample, which corresponds to the number of excited dye molecules, is measured by the transmission (or reflectivity) change of a second ultrashort probe pulse, whose temporal delay with respect to the pump, $\tau_d$, is controlled by a mechanical delay line (Figure 4c,d). In the experiments performed by Quach $et\ al.$,[16] the pump pulse was periodically switched on and off by a mechanical chopper and the corresponding reflectivity of the probe pulse $R_{ON}$ ($R_{OFF}$) was measured, obtaining the differential reflectivity as: $\frac{\Delta R}{R}(\tau_d) = \frac{R_{ON}-R_{OFF}}{R_{OFF}}$. This signal is proportional to the number of excited state molecules, through their ground state bleaching and stimulated emission signals, which in turn is proportional to the energy density stored in the cavity, so that measuring $\Delta R/R(\tau_d)$ as a function of pump-probe delay allows one to track in real time the process of charging and discharging of the microcavity.

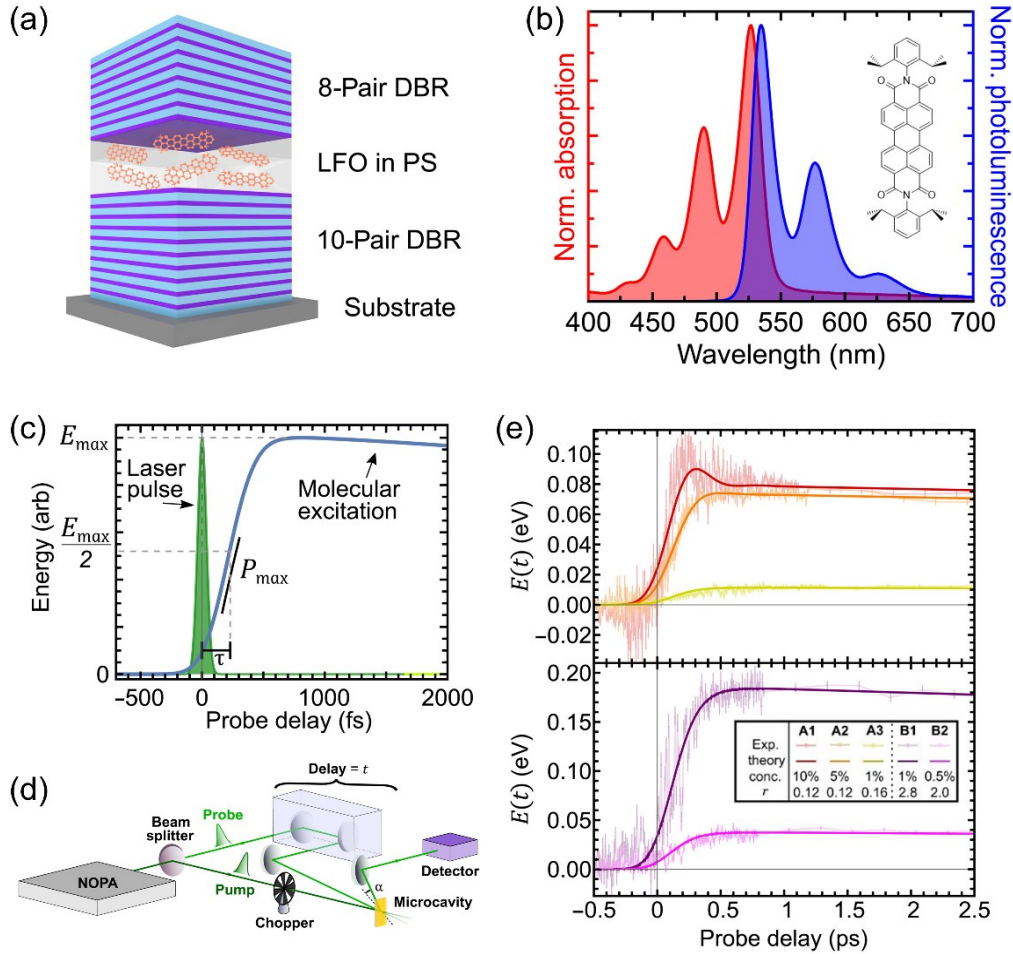

**Figure 4**. (a) Schematics of a microcavity with LFO molecules. (b) Absorption (red line, left vertical scale) and PL (blue line, right vertical scale) spectra of 1% LFO in the PS film. (c) Example of ultrafast transient absorption spectroscopy: A laser pulse (green line) excites the LFO molecules whose energetics are measured by a delayed probe pulse. The measurement allows the rise time ($\tau$), peak energy density ($E_{max}$), and peak charging power ($P_{max}$) to be estimated, as highlighted in the figure. (d) Schematics of the experimental setup used for ultrafast characterization of the sample reflectivity. (e) Time-resolved energy density of the microcavities. The inset shows the concentration of LFO and the photon density ($r$) used for each sample. Reproduced under terms of the CC-BY license.[16] Copyright 2022, The Authors, published by American Association for the Advancement of Science.

Figure 4e compares the temporal dynamics of the energy density in the microcavities (proportional to $\Delta R/R(\tau_d)$) for different loadings, with LFO film concentrations varying from 0.5% to 10%. In all microcavities, one observes a fast rise of the signal followed by a slow decay. By analyzing the data, taking into account the instrumental response function, the authors can demonstrate that, as the number of molecules in the microcavity increases, its charging power density remarkably increases. This superextensive property means that it takes less time to charge a single microcavity containing $N$ molecules than it would to charge $N$ microcavities, each containing a single molecule, even if the latter were charged simultaneously. Furthermore, one microcavity with $N$ molecules would store more energy than $N$ microcavities, each of which contains a single molecule. The experimental data were well reproduced by a numerical model based on the solution of the Lindblad master equations for a collection of $N$ TLS, each with dephasing time $T_2$ and relaxation time $T_1$, coupled to a cavity with light-matter coupling strength and cavity decay rate. The continuous solid lines in Figure 4e mark the results of the model, which are in very good agreement with the experimental data.

While demonstrating superextensive charging, the study by Quach *et al.*[16] did not show controlled storage and discharge of the accumulated energy, since the light energy absorbed by the cavity was re-emitted on an ultrafast timescale. The key challenge for practical applications of organic microcavities as solid-state QBs is the design and realization of devices in which energy can be efficiently stored, and extracted on demand.

To address such challenge, the active material of the cavity can be designed as a donor/acceptor pair, in order to incorporate an ensemble of TLSs (singlet excitons) coupled to the cavity mode for the charging phase and an acceptor organic material with meta-stable states characterized by a longer lifetime to which the excitation of TLSs can be transferred. One possibility is to exploit the triplet states of organic molecules, which have lifetimes ranging from micro to milliseconds or longer.[92-94] Triplet states can be reached either by intersystem crossing (ISC), a nonradiative transition from photogenerated singlet exciton to a triplet exciton induced by spin-orbit coupling, or singlet exciton fission, that is the conversion of a photogenerated singlet exciton into two correlated triplet excitons.[95, 96] To this purpose, different materials might be used, whose properties are summarized in other review papers.[92, 94, 97, 98] An important step towards the implementation of a QB device with donor/acceptor organic molecules has been recently reported by Tibben *et al.*,[99] who designed and realized a Fabry-Pérot microcavity with Ag mirrors and two organic layers, one for the absorption of light and the other for the storage through molecular triplet states. Rhodamine 6G in polyvinyl alcohol (PVA) was chosen as the donor layer (charging), while Palladium tetraphenylporphyrin molecules dispersed in poly (methyl methacrylate) (PMMA), which have efficient ISC to molecular triplet states, constituted the acceptor layer (storage). A self-discharge time of tens of µs was reported, a promising result towards the achievement of a QB.

It is worth mentioning that according to the original proposal of Ferraro *et al.*,[13] the charging and discharging recipes can also be achieved by varying the detuning between the cavity resonance and the exciton transition. Such an approach requires effective methods for tuning the cavity mode wavelength. For organic microcavities, various approaches have been proposed, including, for example, the exploitation of thermo-optic effect,[100] and the use of an electric field to control the alignment of nematic liquid crystals.[101] In particular, the alignment of liquid crystals through electric fields allows for varying the effective refractive index of the cavity active layer incorporating poly(*p*-phenylenevinylene) (PPV) as the organic semiconductor and tuning the cavity mode wavelength by 56 nm. In another approach, the active layer was made of a dye-doped dielectric elastomer, whose thickness was varied by exploiting the Maxwell pressure generated by the electrostatic attraction between the opposite charges of two electrodes deposited on the cavity mirrors, respectively.[102] A shift of the cavity mode wavelength by 40 nm was demonstrated. More recently, all-optical and ultrafast control of the wavelength of the polariton branch in an organic microcavity operating in the strong coupling regime has been reported.[65] In this work, a BODIPY-Br molecule dispersed in a PS matrix was strongly coupled to a microcavity formed by two DBR mirrors. Additionally, a layer of binaphthyl-polyfluorene (BN-PFO), which was not coupled to the cavity, was incorporated. Upon optical pumping with ultrashort laser pulses, the partial bleach of the BN-PFO ground state allows for varying the effective refractive index of the active layer of the cavity and reversibly tuning the energy of the lower polariton branch by 8 meV on sub-picosecond timescales.

Overall, despite the fast exciton decoherence, organic materials present various key features that are promising for the realization of QBs. First, the large exciton binding energy and TLS energy spacing allow for room temperature operation. Microcavities with metal and all-dielectric mirrors achieving $Q$ factors of the order $10^3$ have been realized and both strong and ultrastrong coupling reported.[65, 90] The properties of organic microcavities can be easily tuned by temperature, electric fields and optical beams.[65, 100, 101] Furthermore, it is possible to implement all the three main processes involved in the operation of a QB (charging/superabsorption, storage and discharging) by including different organic materials in the microcavity, one of which supports superabsorption and transfers the energy to a second layer capable of storing it, for example by using long-lived non-emitting states such as charged states or triplets.[99] The discharge step can then be potentially achieved using hole or electron transporting layers. Notably, organic materials with various properties can be synthesized at low cost and processed by deposition methods (spin coating, drop casting and printing technologies) which are easily scalable. In this respect, the synthesis and deposition technologies developed for organic optoelectronic devices[103, 104] are fully available for the development of QBs.

### 3.2. Microcavities with Inorganic Nanostructures

The light-matter coupling in inorganic microcavities with semiconductors nanostructures (such as QWs and QDs) has been investigated intensively in the last thirty years, leading to microcavities operating in various regimes of coupling. For instance, Weisbuch *et al*.[105] reported strong exciton-photon coupling at 5 K in multiple GaAs QWs embedded in a GaAlAs layer, with DBR mirrors made of stacks of GaAlAs/AlAs layers. Rabi splitting typically obtained for inorganic semiconductor microcavities ranges from a few to a few tens of meV. For example, a Rabi splitting of about 9.5 meV was reported in GaAs/AlAs microcavity with six InGaAs QWs,[106] 17.5 meV in Zn-Cd-Se QWs sandwiched between $SiO_2/TiO_2$ DBR,[107] and 50 meV in a $GaN/Al_{0.2}Ga_{0.8}N$ multiple QWs with DBR mirrors.[108] Ultrastrong coupling was also demonstrated for the intersubband transition in a waveguide with multiple GaAs QWs separated by $Al_{0.33}Ga_{0.67}As$ barriers, with the possibility of optically varying the coupling from weak to ultrastrong on ultrafast timescales.[86, 109] A more detailed discussion regarding the various microcavities with inorganic semiconductor nanostructures, the materials used, and the investigation of light-matter coupling and superradiance effects in such devices can be found in recent viewpoint articles and reviews.[110-114]

In the following, we provide a few examples of semiconductor QDs in microcavities, which constitute a potential platform for the experimental implementation of scalable solid-state Dicke QBs, since their employment as qubits has been reported in various works and is being used for quantum technological applications.[115, 116] QDs are nanocrystals with size, $2r$, typically smaller than a few tens of nm, in which electrons are confined in three spatial dimensions and whose bandgap energy is size-dependent.[115, 117-121] When the size of the QD is small compared to the exciton Bohr radius, $r < a_B$, the energy states can be derived by using the particle in a box model and atomic-like states are obtained. A size-dependent bandgap for the QD, $E_{g,QD}(r)$, can be derived:[121] $E_{g,QD}(r) \propto \frac{\hbar^2\pi^2}{2Mr^2}$, where $M$ is the exciton effective mass. QDs with absorption and emission in the whole UV, visible, and near-infrared spectral range can be synthesized, by controllably varying size and composition (Figure 5a,b). Moreover, core-shell structures are designed and realized to improve the fluorescence efficiency and stability.[117] QDs with PL quantum yield approaching unity in solution have been reported.[122] QDs can be made by various methods, including MBE, metal-organic chemical vapor deposition, lithography, and by solution processing.[115] Advantages of colloidal QDs are the possibility to be incorporated in various polymer and inorganic matrices[123-125] and processed by solution methods, whereas lithography and epitaxial fabrication provide a tight control over the size and assembly of QDs.

For the development of Dicke QBs, QDs have to be integrated in microcavities. In the following, some examples of microcavities incorporating single and multiple QDs and operating both in the weak and strong coupling regime are provided, which could constitute the basis for the future implementation of solid-state Dicke QBs. Figure 5c illustrates an example of a Fabry-Pérot microcavity made of two Ag mirrors and an active layer of CdSe/ZnS colloidal QDs dispersed in PMMA.[126] The cavity is made of a bottom 100 nm thick Ag mirror, on top of which a film of QDs/PMMA is deposited by spincoating, while a 40 nm thick Ag film deposited on the QDs/PMMA layer constitutes the top mirror. Microcavities with different wavelengths of the resonant mode were fabricated by varying the thicknesses of the QDs/PMMA film. The coupling of the QDs to the cavity was evidenced by the line narrowing of their PL spectrum compared to a film out of the cavity (Figure 5d). Moreover, the angular distribution of the QDs emission was effectively tailored by the cavity mode, as shown in Figure 5e,f.[126] Strong exciton-photon coupling for colloidal QDs placed in a tunable Fabry-Pérot cavity has been reported by Dovzhenko *et al.*[127] The cavity was made of a bottom flat and a top convex metal mirror, whose alignment and separation were controlled by several micrometric screws and finely tuned by piezoelectric actuators. This configuration of the microcavity allows for enhancing the mode volume. The gap between the mirrors was filled with a solution of an immersion oil containing semiconductor CdSe(core)/ZnS/CdS/ZnS(multishell) QDs. Evidence of the lower polariton branch was found by analyzing the PL properties as a function of the cavity detuning. A coupling strength as large as 154 meV was estimated. In another approach, strong exciton-photon coupling was demonstrated in a high-$Q$ bilayer cavity with CdSe/ZnS core/shell QDs, with coupling strength of the order of 20-30 meV.[128] Xu *et al.*[129] evidenced Rabi flopping and a strong coupling regime for colloidal CdTeSe(core)/ZnS(shell) QDs deposited on a SU8/Si bilayer by dropcasting. Recently, strong coupling has been reported for colloidal CdSe nanoplatelets, with a Rabi splitting of 74-76 meV.[130, 131]

Strong exciton-photon coupling was also reported for a single $In_{0.3}Ga_{0.7}As$ QD in a micropillar.[132] The cavity was realized by sandwiching a GaAs cavity between two AlAs/GaAs DBRs (20 and 23 pairs for the top and bottom mirrors, respectively), while a layer of InGasAs QDs was grown at the antinode of the cavity. All layers were epitaxially grown by MBE. The micropillars were fabricated by electron-beam lithography and reactive ion-etching. Using such an approach, micropillars with diameters of 1-2 µm and $Q$ factors of $10^3$-$10^4$ were realized (Figure 5g). The strong coupling regime was evidenced by controlling the exciton-cavity detuning with temperature, exploiting the temperature dependence of the exciton bandgap energy and refractive index, and observing the characteristic anticrossing behavior (Figure 5h). A vacuum Rabi splitting of about 140 µeV was achieved, thanks to the small mode volume and the high $Q$ factor.

Figure 5i illustrates another example of a micropillar cavity with a single embedded QD and electrical control.[133] The device consisted of a $\lambda$-cavity with two GaAs/$Al_{0.9}Ga_{0.1}$As DBR mirrors grown by MBE (a bottom mirror made of 30 pairs and a top mirror made of 20 pairs). The micropillar cavity (2.9 µm diameter) was defined *in situ* on single QDs by lithography and reactive ion etching. The resulting micropillar was connected to a circular external frame by 4 ridges for electrical contacts. The choice of this configuration combined with p-i-n doping allows for effectively applying a bias to the cavity and finely tuning the QD transition to the cavity mode resonance, as well as to make the charge environment nearby the QD more stable.[133, 134] Figure 5j shows a micro-PL map of the device, highlighting bright emission of the QDs at the pillar center. In such architectures, coherent control of the QDs was demonstrated at the few photon level, while the QDs were efficiently isolated from the solid state environment. Using a similar device, Wenniger *et al.*[135] performed an experimental study of the energy transfer between interacting QDs and light fields. The authors provided a quantitative evaluation of the energy that can be extracted from the quantum device and used to drive another system (a laser field). These results are highly relevant for the experimental implementation of a QB based on microcavities with semiconductor QDs.

Jahnke et al.[136] realized a pillar microcavity with about 200 QDs coupled to the cavity mode. The cavity was made by two AlAs/GaAs DBRs (a top mirror with 20 pairs and a bottom mirror with 23 pairs in $\lambda/4$ configuration) sandwiching a $\lambda$-thick active layer with self-assembled InGaAs/GaAs QDs. In these microresonators operating at 10 K, superradiance was demonstrated by measuring the photon correlations in the light emitted by the devices upon picosecond pulsed excitation. As illustrated in Figure 5k(1)-(4), the photon correlation function $g^{(2)}(\tau=0)$ provides direct evidence of the superradiance, for which $g^{(2)}(\tau=0)>2$ is expected (for coherent emission $g^{(2)}(\tau=0)=1$, while for spontaneous emission $g^{(2)}(\tau=0)$ close to 2 is expected). The properties of the light emitted by the QDs in the micropillars are shown in Figure 5l,m. In particular, the duration of the emitted pulse (about 20 ps) is much shorter than the spontaneous recombination time of the QDs (about 200 ps). Interestingly, $g^{(2)}(\tau=0)>>2$ was observed before and after the maximum emission intensity, especially at the lower excitation intensities.

These findings highlight the existence of superradiance emission and quantum–mechanical correlations in the ensemble of QD emitters incorporated in the micropillar cavity. Overall, these works strongly support these nanostructures as a platform for the experimental realization of Dicke QB. Indeed, most of semiconductor nanostructures have typical TLS energy spacing ranging from hundreds of meV to a few eV, much larger than thermal energy. The low exciton binding energy, however, requires low temperature operation. Cavities made with inorganic semiconductors have high $Q$ factors ($10^3$-$10^4$),[132] which allows achieving the strong and ultrastrong coupling regimes.[86]

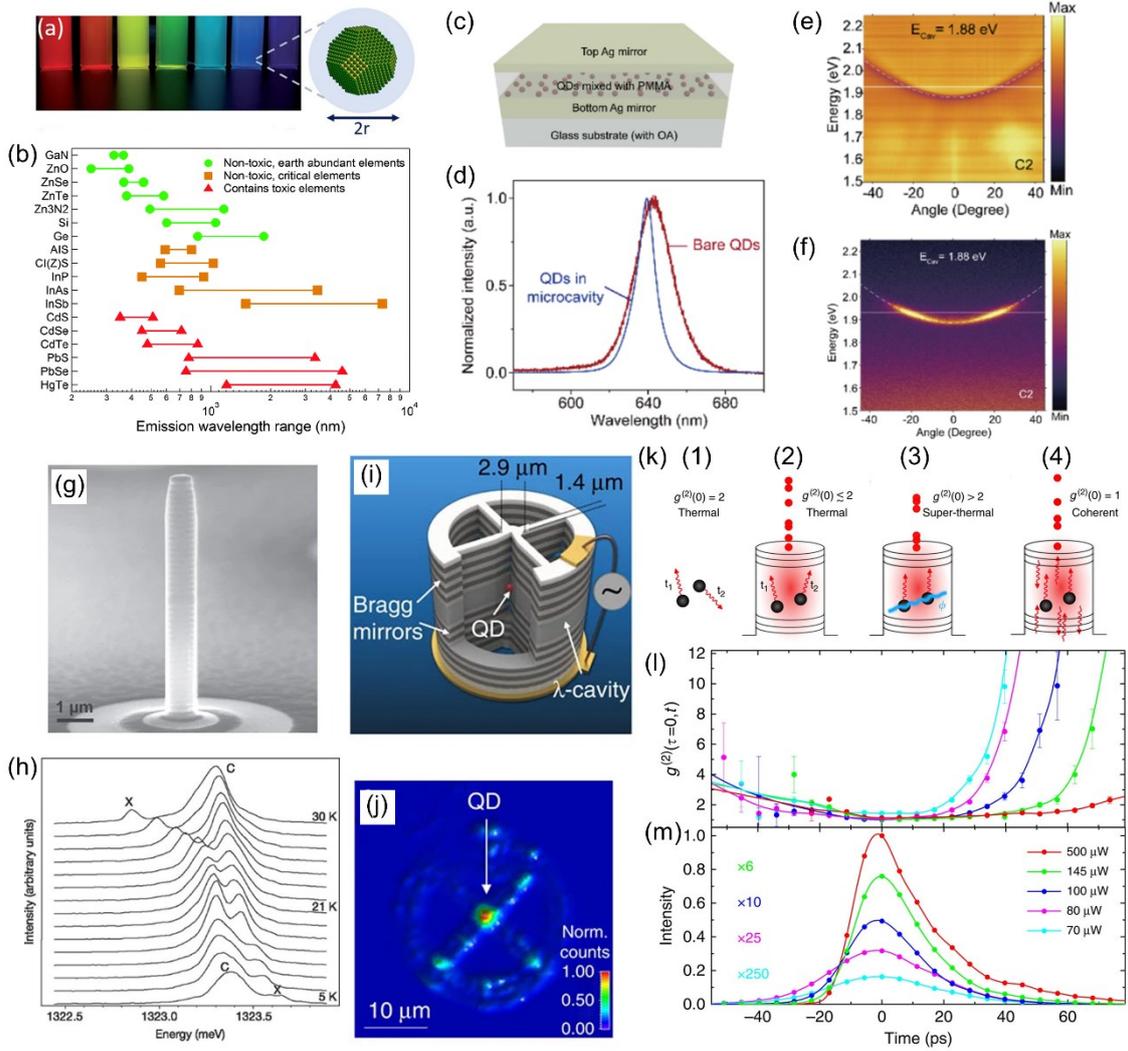

**Figure 5.** (a) Photograph of the light emitted by dispersions of CdSe QDs with a size in the range of 2-6 nm. Reproduced with permission.[118] Copyright 2011, Elsevier. A core shell QD, composed of a crystalline core of size $2r$ and a shell (blue area), is schematized on the right. Reproduced with permission.[115] Copyright 2021, American Chemical Society. (b) Examples of the intervals of emission wavelength tunability of QDs. Reproduced under terms of the CC-BY license.[119] Copyright 2015, The Authors, published by The Royal Society of Chemistry. (c) Sketch of a Fabry-Pérot microcavity with CdSe/ZnS QDs dispersed in PMMA. (d) PL spectra of a PMMA film with CdSe/ZnS QDs out of the cavity (red curve) and in the cavity (blue line). (e),(f) Examples of angle-resolved reflectivity (e) and PL (f) spectra for a cavity with resonant energy at zero degree angle, $E_{cav}$=1.88 eV. Reproduced with permission.[126] Copyright 2022, Optica Publishing Group. (g) Scanning electron micrograph of a micropillar. (h) Variation in temperature of the PL spectra of a micropillar cavity. Reproduced with permission.[132] Copyright 2004, Macmillan Magazines Ltd. (i) Schematics of a microcavity with a single QD located at the center of the sample, as highlighted by the PL map shown in (j). Reproduced under terms of the CC-BY license.[133] Copyright 2016, The Authors, published by Spinger Nature. (k) Illustration of different emission regimes and related properties of the correlation function, $g^{(2)}$ (0). (l),(m) Temporal evolution of the $g^{(2)}$ (τ=0, $t$) (l) and intensity of emission intensity (m) upon pulsed excitation of the micropillar cavity with various intensities, as indicated in the inset of (m). Reproduced under terms of the CC-BY license.[136] Copyright 2016, The Authors, published by Spinger Nature.

Notably, semiconductor nanostructures and QDs exhibit properties that are highly tunable by external fields,[137, 138] a property used to tune the light-matter interaction.[86, 139] This property can be potentially exploited for the experimental implementation of controlled charging/discharging protocols in QBs. Similarly to organic materials, energy transfer to long-lived states, as for example triplets in graphene QDs,[140] might be explored for the storage phase. Recent advances toward the large scale production of QDs[141-143] are also relevant for the realization of QBs.

### 3.3. Microcavities with Perovskites

The interest in lead halide perovskites as a class of semiconductors for studying light-matter interactions in microcavities has grown significantly in recent years, due to their high oscillator strength, strong exciton binding energy, high PL quantum yield, and widely tunable bandgaps.[144-146] Metal halide perovskites typically exhibit an $ABX_3$ structural formula (where A is a cation, B is a metal, and X is a halide) and possess a crystal structure illustrated in Figure 6a.[144] The high structural design flexibility, ranging from purely inorganic to hybrid inorganic-organic composition, and from 3D networks to 2-dimensional (2D) layered materials and lower dimensional materials (such as nanowires and QDs), combined with the potential for solution processing, has made halide perovskite highly suitable for optoelectronic applications.[144, 147-149] In particular, 2D Ruddlesden–Popper layered hybrid perovskites[148, 150, 151] feature a multiple QW structure, where the inorganic layers are sandwiched between organic cations (Figure 6b), making them appealing for investigating exciton-photon coupling at room temperature.[152, 153]

Figure 6c shows one of the first planar microcavities realized with a 2D layered perovskite [$(C_6H_5C_2H_4–NH_3)_2PbI_4$, chemical structure shown in Figure 6d], which exhibits an exciton transition energy at 2.4 eV (Figure 6e).[154] The microcavity comprised a bottom dielectric mirror, while a thin film of $(C_6H_5C_2H_4–NH_3)_2PbI_4$ and a PMMA spacer were deposited by spin-coating. A top Ag mirror was deposited on the PMMA layer via thermal evaporation. A Rabi splitting of 150 meV was determined by angle-resolved reflectivity measurements, demonstrating the characteristic anticrossing behavior between the upper (UPB) and lower (LPB) polariton branches (Figure 6f). Strong coupling was also demonstrated for a $(C_6H_5C_2H_4–NH_3)_2PbCl_4$ 2D perovskite with absorption and emission in the ultraviolet (3.6 eV).[155] A Rabi splitting of 230 meV was obtained. Wang *et al.*[156] realized a microcavity with a $SiO_2/Ta_2O_5$ bottom and top DBR mirrors and a 2D organic-inorganic perovskite crystal exfoliated and transferred to the bottom DBR using the Scotch tape method. In another study, a Rabi splitting of about 110 meV and a $Q$ factor >1000 were reported for a microcavity with a single crystal of phenethylammonium lead iodide perovskite.[157] Recently, Laitz *et al.*[158] obtained a Rabi splitting of 260 meV in a wedged microcavity with phenethylammonium lead iodide perovskite $(C_6H_5(CH_2)_2NH_3)_2PbI_4$ (PEA$_2$PbI$_4$).

Strong coupling has also been evidenced for 3D perovskites.[159, 160] Figure 6g shows an example of microcavity with a 3D bromide hybrid perovskite, $CH_3NH_3PbBr_3$, (MAPB) active layer. The microcavity was completed by a bottom DBR mirror and a top Ag mirror. The absorption and PL properties of the $CH_3NH_3PbBr_3$ film are shown in Figure 6g(ii). The microcavity has a $10^2$ $Q$ factor, with a Rabi splitting up to 70 meV (Figure 6h).[160]

More recently, strong coupling of $CsPbBr_3$ QDs embedded in a metallic microcavity has been reported, with a Rabi splitting of 87 meV at room temperature.[161] This result is especially interesting given the possibility of synthesizing $CsPbBr_3$ QDs with monodispersed and tunable size, and emission properties in the blue-green spectral range on substrates.[162] Notably, Bertucci *et al.*[163] recently reported a microcavity made entirely by solution processing. The $CsPbBr_3$ QDs were embedded between two DBRs of silica and titania that were realized by sol-gel deposition. The microcavity exibits a $Q$ factor of 220.

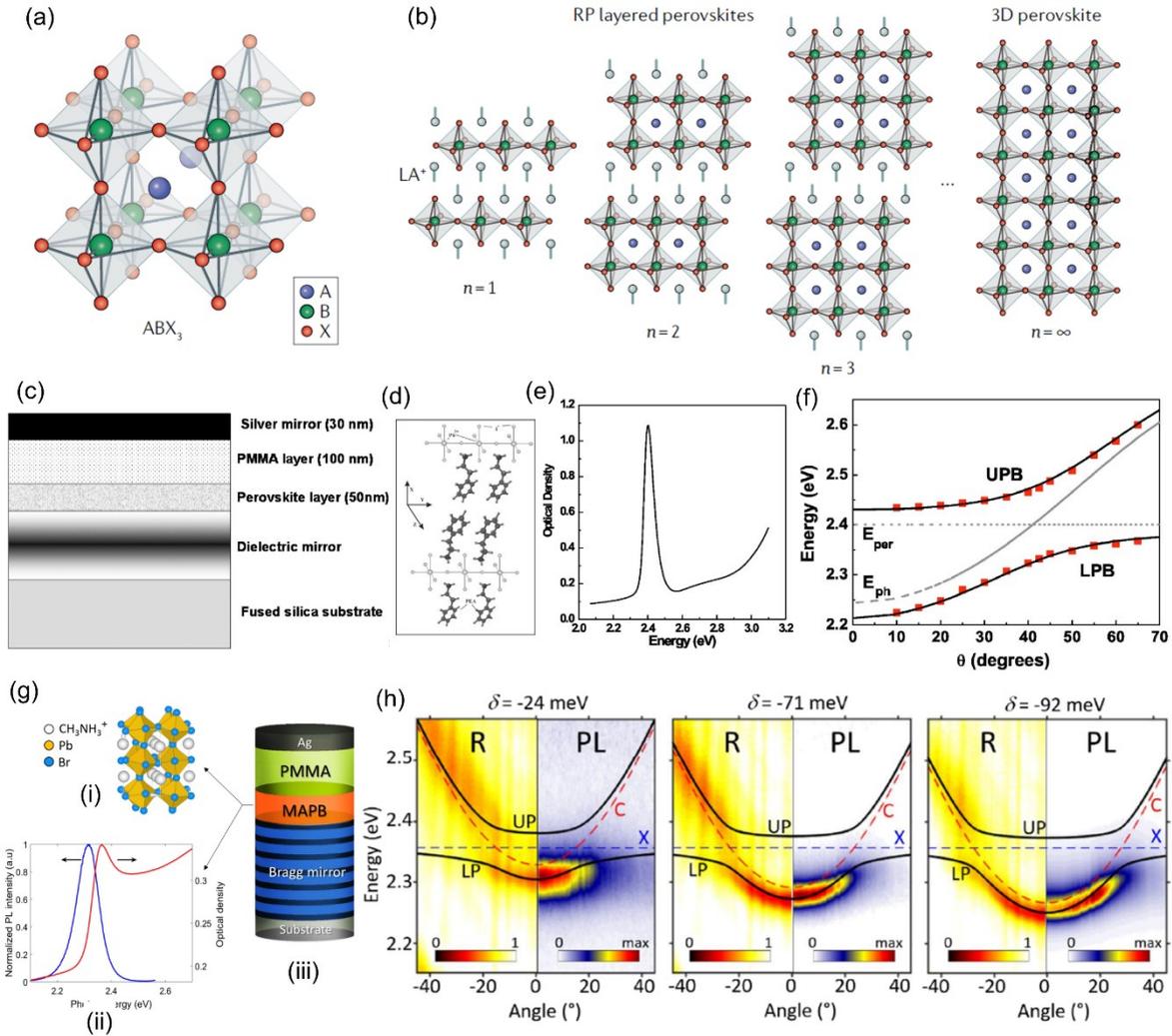

**Figure 6**. (a),(b) Scheme of the structure of a 3D perovskite with a general formula ABX$_3$ (a) and of Ruddlesden–Popper (RP) layered perovskites (b). Reproduced with permission.[144] Copyright 2019, Springer Nature Limited. (c) Sketch of a Fabry-Pérot microcavity with a 2D (C$_6$H$_5$C$_2$H$_4$–NH$_3$)$_2$PbI$_4$ perovskite layer, whose structure is shown in (d). (e) Absorption spectrum of a thin film of (C$_6$H$_5$C$_2$H$_4$–NH$_3$)$_2$PbI$_4$. (f) Angular dispersion of the UPB and LPB as obtained by angle-resolved reflectivity data (red symbols). The continuous black lines are fits to the data by a two-level model. The horizontal dotted line highlights the energy of the uncoupled perovskite exciton, whereas the dashed line shows the angular dispersion of the bare cavity mode. Reproduced with permission.[154] Copyright 2006, American Institute of Physics. (g) Scheme of the structure of the 3D MAPB perovskite (i). (ii) Absorption and emission spectra of MAPB. (iii) Schematics of a microcavity. (h) Angle-resolved reflectivity (left panels) and PL (right panels) spectra measured for microcavities with three different detunings, $\delta$, between the cavity mode and exciton energy. Reproduced with permission.[160] Copyright 2019, American Chemical Society.

The aforementioned examples show the great potential that perovskite materials hold for the implementation of QBs. Their TLS energy spacing of order of few eV allows for room temperature operation, as supported by the observation of strong coupling at room temperature in microcavities with $Q$ factors of 10-10$^2$.[154,160] Moreover, the properties of perovskite materials can be tuned by external fields, such as electrical field and optical pulses, also on ultrafast timescales.[164, 165] The structural design flexibility of perovskite materials allows for the tuning of their properties, enabling the creation of materials with long-lived states[166, 167] that can be utilized in the energy storage phase of QBs. Photoelectric conversion effects observed in perovskite materials[168] could also be potentially harnessed for the discharging phase. Importantly, recent advances in large-scale synthesis and processing of perovskite materials, driven by the application in solar cells,[169, 170] are highly relevant for future upscaling of potential QB production.

### 3.4. Superconductors

In recent years, various superconducting circuits have been designed and manufactured to realize quantum states, study collective effects, and simulate Hamiltonians relevant to quantum optics.[171-174] An experimental platform employs superconducting qubits, known as transmons (Figure 7a,b),[175] as the TLSs, embedded in a microwave cavity, which is a coplanar waveguide resonator.[171] Figure 7c,d shows optical micrographs of a device with three superconducting qubits (highlighted as "A", "B" and "C") coupled to the resonator, where a clear Rabi mode splitting is observed (Figure 7e). This device was used for the simulation of the Tavis-Cummings[176] model.[171] In another study, collective modes of qubits were evidenced in a device with 20 transmons coupled to a resonator.[173] In these devices, the resonator is typically fabricated using optical lithography, etching processes, and metal evaporation, while electron beam lithography and evaporation of metals such as aluminum, niobium, or tantalum, along with their oxides, are used to fabricate the transmons.[171, 177, 178] The quality of surfaces, interfaces, and the presence of contaminants can limit qubit lifetime and coherence time. This has driven research efforts to identify the most suitable materials and interfacial properties. For instance, Place *et al.*[177] reported an increase in transmon lifetime by replacing niobium, commonly used for transmon fabrication, with tantalum. Other studies have shown that native niobium oxide can cause microwave loss.[179] Preventing the formation of this oxide by encapsulating the niobium layer with different materials has significantly enhanced the qubit relaxation time.[178]

Building on the significant results demonstrated with the aforementioned superconducting qubits, transmons coupled to microwave resonators appear to be a natural platform for the implementation of Dicke QBs targeting applications in quantum technologies. A first preliminary step toward the fabrication of QBs based on superconducting circuits has been made by the authors of Ref. [17], who fabricated devices containing a single transmon qutrit (i.e, a three levels system)[17, 19] in a superconducting resonator. Although the use of a single qutrit does not allow for investigating collective effects, these works constitute a first important step toward the experimental fabrication of Dicke QBs based on superconductors. The realization of devices with 20 transmons coupled to a resonator[173] is highly promising for investigating collective effects in Dicke QBs based on superconducting circuits. Although these devices need to be operated at low temperature, they features long coherence lifetime (order of 1-100 μs) suitable for testing charging/storing/discharging processes.[180] Important steps towards the large-scale fabrication of superconducting circuits are also being developed.[181] Superconducting QBs can also be simulated, with some limitations, on platforms like the IBM Quantum Experience.[182, 183]

It is worth mentioning that superconductor circuits have been also considered for the potential realization of bosonic QBs with non-linear interactions.[60] In particular, Andolina *et al.*[60] considered a circuit made by two superconducting LC resonators that are coupled by a Josephson junction[184-186] and demonstrated that the interaction Hamiltonian can be non-linear in some regimes.

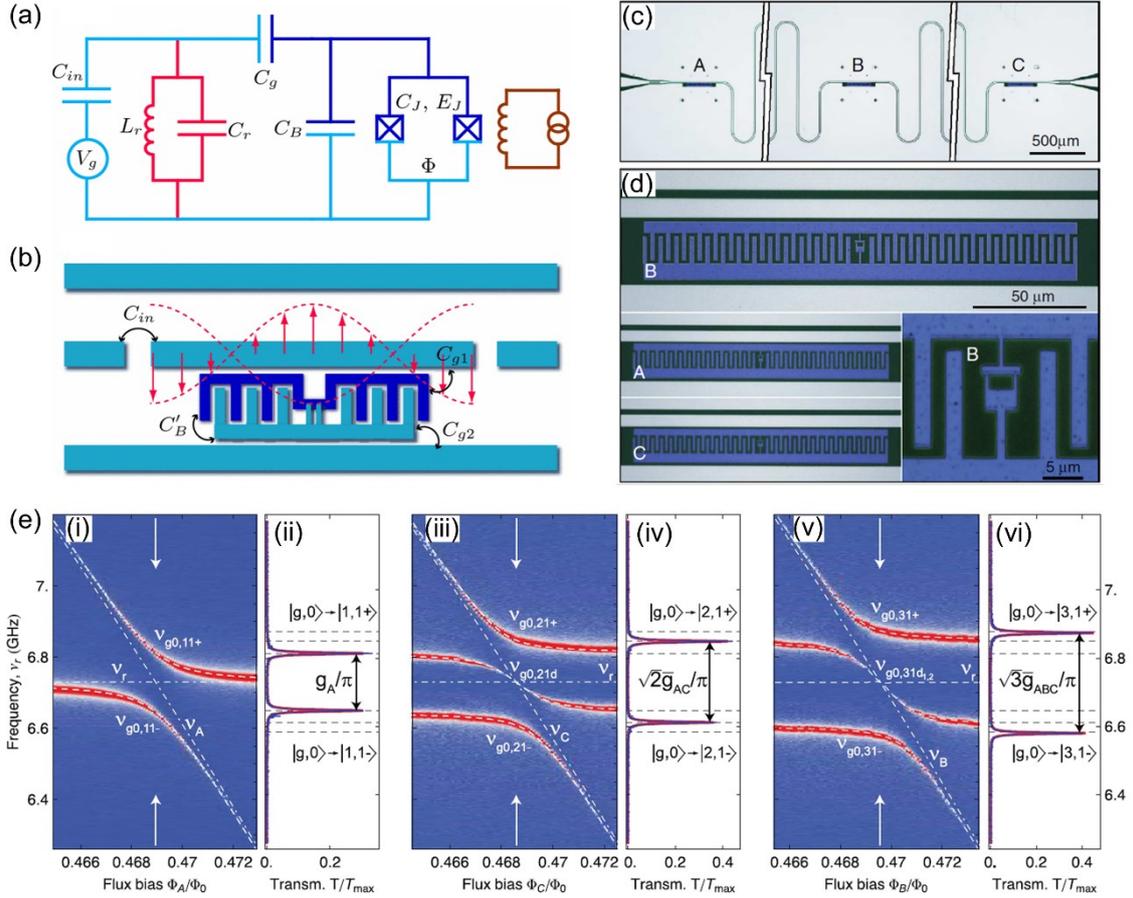

**Figure 7**. (a),(b) Effective circuit diagram (a) and schematic illustration of a transmon device. $C_{in}$, $C_r$, $C_g$, $C_B$, $C_j$, $C_{g1}$, $C_{g2}$, $C_B'$ are capacitances. $E_J$ indicates the Josephson energy. Reproduced with permission.[175] Copyright 2007, The American Physical Society. (c) Optical micrographs of the microwave resonator with three qubits indicated as A, B, and C. The coplanar microwave waveguide is shown truncated. (d) Magnified optical image of the transmon qbits. Inset: Magnified view of the SQUID loop of qubit B. (e) Dependence of the transmission spectrum, here $T/T_{max}$, of the resonator as a function of the external flux bias. Data shown in (i) and (ii) are obtained by varying the external flux of qubit A (i). The resonant transmission spectrum at degeneracy [highlighted by vertical arrows in (i)] is shown in (ii). Data shown in (iii) and (iv) are obtained by varying the external flux of qubit C while A has the same resonator frequency and B is far detuned. The data shown in (v),(vi) are obtained by varying the external flux of qubit B, while A and C are at degeneracy. Reproduced with permission.[171] Copyright 2009, The American Physical Society.

Table 1 summarizes the key properties of the materials discussed in the previous Sections, and that can be potentially used as TLS coupled to microcavities for the realization of QBs.

**Table 1.** Properties of TLSs coupled to microcavities realized with different materials.

| Material | TLS energy spacing [eV] | Cavity quality factor | Maximum Rabi splitting [meV] | Strong coupling [Y/N] | Ultra-Strong coupling [Y/N] | Cavity tunability [Y/N] | Ref. |
|---|---|---|---|---|---|---|---|
| LFO | 2.36 | - | 100 | Y | N | N | [16] |
| TDAF | 3.5 | ~30 | ~1000 | Y | Y | N | [90] |
| BODIPY-Br/ BN-PFO | 2.3 | ~2600 | 103 | Y | N | Y | [65] |
| CdSe nanoplatelets | 2.4 | 60 | 74.6 | Y | N | N | [131] |
| GaAs QWs | Hundreds meV | - | Tens of meV | Y | Y | Y | [86, 109] |
| $In_{0.3}Ga_{0.7}As$ QD | 1.3 | $10^3$-$10^4$ | 0.14 | Y | - | - | [132] |
| $(C_6H_5C_2H_4-NH_3)_2PbI_4$ | 2.4 | 25 | 150 | Y | - | - | [154] |
| $CH_3NH_3PbBr_3$ | 2.3 | $10^2$ | 70 | Y | - | - | [160] |
| $Al/AlO_x/Al$ transmons | ~ A few GHz | (not stated in Ref. but typically between $10^4$ and $10^6$) | ~ 80 MHz | Y | - | - | [171] |

## 3.5 Spin Arrays

Arrays of spins constitute another interesting platform for the experimental realization of QBs. A first step in this direction was reported by Joshi *et al.*,[18] by using nuclear spins of various sizes in star shaped molecules.[187] The authors used different molecules: acetonitrile, trimethyl phosphite, tetramethyl silane, hexamethylphosphoramide, and tetrakis(trimethylsilyl)silane (TTMS). These systems have a common structure, that is a single nuclear spin located on a central atom ($^{13}C$, $^{31}P$, $^{29}Si$, $^{31}P$ and $^{29}Si$, respectively), constituting the battery, interacting with *N* surrounding ancillary spins ($^1H$, *N*=3, 9, 12, 18, 36, respectively), which constitute the charger. Through nuclear magnetic resonance (NMR) methods, the authors can investigate charging schemes, estimate the ergotropy and achieve asymptotic charging. Interestingly, for TTMS molecules, the single spin battery can store the energy for up to 2 minutes before transferring the accumulated energy to another load spin. Although based on a single spin, this work constitutes a first step towards the implementation of QBs based on spin arrays, and allows some charging and discharging properties to be studied at room temperature.

Other recent studies on spin qubits and sensors developed in the framework of quantum technologies[188-191] are also interesting for the implementation of spin array QBs. These studies are basically motivated by the long coherence times and the possibility to build arrays with multiple spins, both properties that are also interesting for implementing the storage phase of QBs and investigating charging speed-up in many body systems. Examples include point defects in bulk solid-state materials and 2D materials[190, 192] such as nitrogen-vacancies (NVs) in diamond,[188, 193] and semiconductor nanostructures.[189, 191]

## 3.6 Perspectives on the Potential Use of Strange Metals for QBs

As introduced in Section 2, in 2020, Rossini *et al.*[20] and Rosa *et al.*[194] proposed, on purely theoretical grounds, a quantum many-body battery model based on the so-called SYK model.[48-50] To the best of our knowledge, this is the only model displaying a genuine quantum advantage. Proposals to realize the SYK Hamiltonian in the laboratory have been put forward, relying on ultra-cold atoms[51] and solid-state systems,[195] such as topological superconductors[52, 53] and graphene QDs with irregular boundaries in strong applied magnetic fields.[54-56] Experimental evidence for the achievement of a regime of strong correlations in the latter systems has been recently reported by Anderson *et al.*,[196] who presented data for the thermoelectric power of these QDs exhibiting strong departures from the Mott formula in high magnetic fields (on the order of 10 Tesla) and elevated temperatures ($T \geq 10$ K). These data are compatible with the emergence of SYK-type correlations.[197] This is clearly a milestone result, paving the way for the potential fabrication of QBs based on arrays of graphene QDs with disordered edges.

We further note that SYK and similar models with random spatial couplings have been theoretically linked to the emergence of "strange metal" behavior in solids (see, for example, Ref. [57]). Strange metals, common in quantum materials (such as twisted bilayer graphene,[59] cuprates,[58] and other strongly correlated systems, such as iron pnictides,[198] heavy fermion metals,[199, 200] infinite-layer nickelates[201]), are metallic phases of matter defying the paradigm of Fermi liquids with "quasiparticles".[202, 203] In Ref. [57], the authors showed that many-body models with random Yukawa interactions inspired by the SYK Hamiltonian lead to strange metal behavior, including a linear $T$-dependent resistivity. Cuprates, being the prototypical materials for this novel state of matter, have been the most extensively studied. In the following, we provide a brief overview of the unique characteristics of strange metals, highlighting the differences from conventional metallicity, with an emphasis on the phenomenology of cuprate high-temperature superconductors (HTSs).

Cuprates have a layered structure with planes containing single copper atoms in the center of a square of oxygen ligands (Figure 8a). The quasi-2D $CuO_2$ planes host strong in-plane correlations, which give rise to intricate correlated quantum phases intertwined with charge, spin, and orbital orders. The immense complexity of cuprates physics is encompassed in the phase diagram which highlights how the properties evolve with doping and temperature. For an in-depth description of their phase diagram, we refer the reader to extensive reviews.[58, 204] Briefly, at high temperatures, the material evolves from an antiferromagnetic (AF) Mott insulator at low doping to a rather conventional metal in the overdoped regime, with a characteristic Fermi-liquid temperature dependence of the resistivity, $\rho \propto T^2$. In between these two regimes, one finds the "strange metal" phase with $T$-linear resistivity, $\rho \propto T$.[58] In this temperature range, quasiparticles, the main building block of a weakly interacting Fermi liquid, cease to exist. The absence of conventional quasiparticles has been confirmed experimentally by several ARPES experiments.[205, 206] The failure of the quasiparticle picture is supported by the linear temperature dependence of resistivity, as we will discuss below. Figure 8b shows one of such measurements performed in the Lombardi group on the HTS $YBa_2Cu_3O_{7-x}$ at optimal doping. In another compound, i.e. $La_{2-x}Sr_xCuO_4$ (LSCO), the linearity of the resistivity vs. temperature has been reported in a broader range of temperatures, exceeding 500 K (Figure 8c).[207-209] These measurements reveal some remarkable features. At high temperatures, the resistivity becomes so large that, in the context of conventional billiard-ball-like quasiparticle transport,[210] the distance a quasiparticle travels between scattering events (which relax momentum) would need to be much shorter than the lattice constant. This scenario, known as a violation of the Mott-Ioffe-Regel (MIR) limit, is not a reasonable assumption.

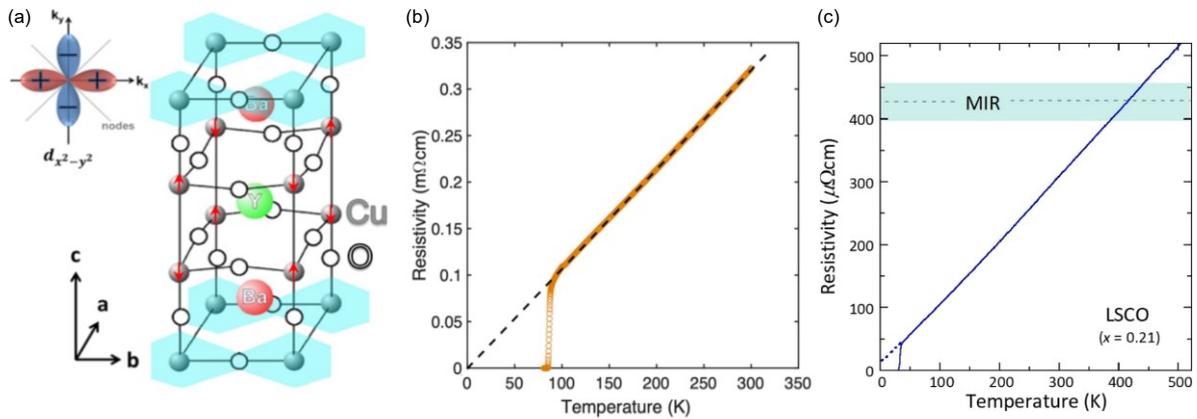

**Figure 8**. a) $YBa_2Cu_3O_7$ unit cell. Light blue regions highlight the presence of CuO chains. The $CuO_2$ planes are characterized by an antiferromagnetic order. The inset shows the superconducting order parameter, $d_{x^2-y^2}$ b) $T$-linear resistivity of a $YBa_2Cu_3O_{7-x}$ film grown at the Chalmers University of Technology and measured close to optimal doping. Note the resistivity intercepts at zero temperature, indicating negligible scattering off static impurities. c) $T$-linear resistivity of a $La_{2-x}Sr_xCuO_4$ ($x$=0.21). The shaded area highlights the MIR boundary, which occurs when the mean free path becomes comparable to the Fermi wavelength. Reproduced with permission.[209] Copyright 2022, The American Association for the Advancement of Science. Reproduced with permission.[208] Copyright 2011, The Royal Society.

This already rules out conventional quasiparticles as charge carriers. In addition, the resistivity is perfectly linear over a very wide temperature range, starting from very high values (above room temperature[207, 208]) and extending down to the temperature scale at which the material becomes superconducting. In a quasiparticle framework, linear temperature scaling can be connected to a specific microscopic mechanism, but only within a narrow temperature range. For instance, electron-phonon scattering can produce the desired linear dependence; however, phonons significantly contribute to scattering only at temperatures above the Bloch-Gruneisen temperature. At lower temperatures, this mechanism is expected to be negligible. Another striking feature is that the linear temperature dependence can be extrapolated to zero temperature (black dashed line in Figure 8b), a region that can be experimentally accessed by suppressing superconductivity in a strong magnetic field.[207, 211] All these anomalies in the $T$-linear resistivity seem to indicate that the momentum relaxation time $\tau_{tr}$ in the system is completely independent of the material's parameters. In the simple case of an isotropic Fermi surface, the connection between the resistivity $\rho$ and $\tau_{tr}$ is given by the Drude formula, $\rho = \frac{m^*}{ne^2\tau_{tr}}$ where $n$ is the carrier density, $m^*$ is the effective mass. Recent theoretical developments[203, 212] have connected the momentum relaxation time $\tau_{tr}$ to the "Planckian time" ($\tau_{Pl} = \alpha \frac{\hbar}{k_B T}$), which involves only the Planck constant $\hbar$ and Boltzmann constant $k_B$ as well as the temperature $T$, but none of the material parameters. A stringent test on whether the time Planckian scale is at play in the cuprates or not is given by the estimate of the proportionality factor, $\alpha$, which needs to be of order unity. For the cuprate families, where the carrier density and the effective mass can be assessed by independent measurements, it has been found that indeed $\alpha$ is on the order of 1.[211] However, while $T$-linearity and Planckian dissipation seem nearly synonymous in hole-doped cuprates, this relationship is not universal. For instance, in some heavy fermion systems, $\tau_{tr}$ significantly deviates from the Planckian limit, and in electron-doped cuprates, this connection has been called into question.[209]

Overall, the fact that the relaxation time in cuprates is independent on the defects and material properties suggests that a fundamentally new transport mechanism might be at play. But what is this mechanism? Recent theories have put forward the so-called "holographic duality" by describing the physics of strongly interacting systems in terms of virtual black holes.[213, 214] The holographic duality has naturally embedded the Planckian timescale through the dynamics of virtual black holes. However, this duality also gives sharp predictions about a transport regime dominated by hydrodynamics[212] i.e., insensitivity to impurities as it is found in the strange metal phase.

The hydrodynamic description for current transport can be achieved in systems that are described by quasiparticles, if charge carriers collide so frequently with each other that the lengthscale associated with momentum-conserving collisions (such as electron-electron scattering) is smaller than both the sample size $W$ and the momentum relaxation length $l_{MR}$. Such a hydrodynamic transport regime is nowadays reached in ultraclean graphene samples[210, 215-218] and other ultra-clean materials such as GaAs QWs[219, 220] where in a certain range of temperature the $e$-$e$ scattering length becomes the shortest length in the system. But this is not the case for cuprates, which are notoriously disordered systems. A possible path to hydrodynamic flow in these materials is offered by a strong entanglement between charge carriers, which can lead to very fast thermalization times. Due to entanglement, carriers lose their locality (breakdown of the quasiparticle picture) and transport is dominated by hydrodynamics. If this is realized, one should be able to observe a viscous flow in cuprates systems with W smaller than $l_{MR}$. This length is given by $\sim v_F \tau_{tr}$, where $v_F$ is the Fermi velocity. Using typical values for YBa$_2$Cu$_3$O$_7$, i.e., $v_F \sim 3 \times 10^5$ m s$^{-1}$ and a temperature of 100 K (slightly above the superconducting transition temperature), one would require device dimensions on the order of 30 nm to enter the hydrodynamic transport regime. This value is within the reach of cuprate nanotechnology,[221] paving the way for an experimental verification of hydrodynamic transport in these complex materials.

A recent work has used a fundamental concept in quantum metrology, the quantum Fisher information, to quantify the level of entanglement of many-body systems by the measurement of a dynamic susceptibility that can for example be derived from inelastic neutron scattering experiments. This technique has been applied to the strange metal phase of a heavy fermion compound, evidencing the strongest entanglement detected to date in any many-body quantum system.[222] Whether such massive entanglement is a general property of the strange metal phase is an important question that should be addressed by future experiments across the strange metal platforms that can be addressed by this novel technique and by the detection of hydrodynamic transport.

Overall, these studies of strange metallicity in cuprates and, more recently, in twisted bilayer graphene near the magic angle[59] suggest interesting perspectives for the experimental realization of QBs based on strongly correlated electron systems.

## 4. Outlook and Conclusion

The Sections above have reviewed materials with the greatest potential for the experimental realization of fully operational QBs. Table 2 summarizes some key features of the considered materials and the related processing methods.[223, 224] Although a variety of materials and microcavity designs can be potentially used for QB fabrication, several challenges and issues must be addressed. One primary issue is related with the long-term storage of the absorbed energy, and the use of the stored energy on demand. For organic materials, an emerging strategy to address this challenge is the use of donor/acceptor systems, where energy absorption and storage are assigned to materials with different properties. To assess the viability of this approach, theoretical and experimental efforts should fully explore how donor/acceptor systems behave when coupled to a common cavity mode, going beyond the models that consider only one type of TLS coupled to the cavity mode. Recent results regarding energy transfer in donor/acceptor systems in microcavities and the role of polariton transitions[225, 226] might be inspiring in this regard. Another interesting proposal for extending the storage time considers the possibility of storing the energy through symmetry-protected dark states.[28]

Recent studies have highlighted the role of material interfaces in limiting the TLS lifetime, prompting research efforts focused on strategies to control and improve TLS lifetime and coherence. These strategies include selecting specific macromolecular systems and multilayers,[16, 99] and controlling the interfaces between the various materials used.[177, 178] Additional studies on superabsorption effects in microcavities by using other organic, inorganic and perovskite materials are also necessary to understand the charging processes in QBs.

Furthermore, understanding which applications can truly benefit from QBs is crucial for directing research efforts towards a specific material platform. Quantum technologies will likely be primary users of QBs, particularly for operating quantum devices that require coherence and entanglement.

To summarize, since their introduction in 2013[9] significant progress has been made in understanding the fundamental properties and key features of QBs. Initial experimental implementations have suggested promising routes for realizing QB devices with superextensive charging capabilities. The demonstration of a fully operational QB requires the convergence of research efforts from various materials science communities and a strong synergy between theoretical and experimental research. Interdisciplinary collaboration between different materials scientists is essential to promote the development of new systems and device structures, and accelerate the realization of quantum batteries. Driven by the need for novel and more efficient energy storage devices specifically addressing quantum technologies, research in QBs can deepen our understanding of light-matter interaction, especially in complex multimaterial systems, and lead to unexpected device configurations.

**Table 2**. Properties of the materials considered for potential QBs realization and the related processing methods.

| Material | Stability | Costs of raw materials[a] | Manufacturing technology of devices | Scalability of synthesis/processing technologies Y/N | Operational temperature (K) |
|---|---|---|---|---|---|
| Metals for mirrors | High | 1-10 Euro/g[b] | Thermal evaporation, electron beam evaporation, sputter deposition | Y | RT[c] |
| Dielectrics for DBR | High | $10^{-1}$-1 Euro/g[d] | Electron beam evaporation, sputter deposition, molecular beam epitaxy | Y | RT |
| Organic molecules | Good. Could be enhanced by suitable device encapsulation[224] | $10$-$10^4$ Euro/g | Spin coating, drop casting, thermal evaporation, blade coating, ink-jet printing | Y | RT |
| QDs | High | $10^3$-$10^4$ Euro/g[e] | Spin coating, drop casting, metal-organic chemical vapor deposition, lithography, ink-jet printing | Y | From few K to RT |
| Perovskites | Good. Could be enhanced by passivation and encapsulation methods[225] | $10$-$10^3$ Euro/g | Spin coating, thermal annealing, exfoliation, antisolvent vapor-assisted crystallization. | Y | RT |
| Normal Superconductors | High[181] | 1-10 Euro/g | Optical lithography, electron beam lithography etching processes, metal evaporation | Y | 10-50 mK |
| High-temperature superconductors | High | $10^2$-$10^3$ Euro/g | Electron-beam lithography, ion beam etching | Y | - |

[a] Costs might differ depending on the purity of the material. [b] Costs estimated for Ag; [c] RT: room temperature; [d] Costs estimated for $TiO_2$ and $SiO_2$; [e] Costs estimated for colloidal QDs.

**Acknowledgements**

F. L. acknowledges the support by the 2D-Tech VINNOVA Competence Center (Grant No. 2019-00068), and the Swedish Research Council under the Projects No 2022-04334. F. L.

would like to thank R. Arpaia and E. Wahlberg for help with the $YBa_2Cu_3O_{7-x}$ resistivity measurements and pictures, and T. Bauch, U. Gran and J. Hofmann for useful discussions. G. C. acknowledges funding from the European Horizon EIC Pathfinder Open programme under grant agreement no. 101130384 (QUONDENSATE). G. C. acknowledges financial support by the European Union's NextGenerationEU Programme with the I-PHOQS Infrastructure [IR0000016, ID D2B8D520, CUP B53C22001750006] "Integrated Infrastructure Initiative in Photonic and Quantum Sciences". M.P. was supported by the PNRR MUR project PE0000023-NQSTI (Italy). M.P. wishes to thank all his collaborators on the topic of quantum batteries and, in particular, G. M. Andolina, D. Rossini, and V. Giovannetti for many useful discussions.

**Conflict of Interest**

M.P. is the co-founder and CSO of Planckian srl.